\documentclass[]{spie}  

\newcommand{\commentout}[1]{}
 
\usepackage{amsmath,amsfonts,amssymb}
\usepackage{graphicx}
\usepackage[colorlinks=true, allcolors=blue]{hyperref}

\title{First measurements and upgrade plans of the MAGIC intensity interferometer}

\author[a]{Juan Cortina}
\author[b]{V. A. Acciari}
\author[c]{A. Biland}
\author[b]{E. Colombo}
\author[a]{C. da Costa}
\author[a]{C. Delgado}
\author[a]{C. D\'{\i}az}
\author[d]{M. Fiori}
\author[e]{D. Fink}
\author[a]{T. Hassan}
\author[a]{I. Jim\'enez-Mart\'{\i}nez}
\author[f]{E. Lyard}
\author[d]{M. Mariotti}
\author[a]{G. Mart\'{\i}nez}
\author[e]{R. Mirzoyan}
\author[d]{G. Naletto}
\author[a]{M. Polo}
\author[f]{N. Produit}
\author[a]{J. J. Rodr\'{\i}guez}
\author[e]{T. Schweizer}
\author[f]{R. Walter}
\author[g]{C.~W. Wunderlich}
\author[d]{L. Zampieri}
\author[]{the MAGIC and LST collaborations}

\affil[a]{CIEMAT, Avda. Complutense, 40, Madrid, Spain}
\affil[b]{IAC, E-38200 La Laguna, Spain}
\affil[c]{ETH, CH-8093 Zurich, Switzerland}
\affil[d]{Universit\`a di Padova, I-35131 Padova, Italy}
\affil[e]{Max-Planck-Institut f\"ur Physik, D-80805, Munich, Germany}
\affil[f]{Observatoire Astronomique de Gen\`eve, CH-1290 Versoix, Switzerland}
\affil[g]{Universit\`a di Pisa and INFN Pisa, I-56126 Pisa, Italy}

\authorinfo{Further author information: (Send correspondence to J.C.)\\J.C.: E-mail: juan.cortina@ciemat.es}

\begin{document}
\maketitle

\begin{abstract}
  The two MAGIC 17-m diameter Imaging Atmospheric Cherenkov Telescopes have been equipped to work also as an intensity interferometer with a deadtime-free, 4-channel, GPU-based, real-time correlator. Operating with baselines between $\sim$40 and 90 m the MAGIC interferometer is able to measure stellar diameters of 0.5 - 1 mas in the 400-440 nm wavelength range with a sensitivity roughly 10 times better than that achieved in the 1970’s by the Narrabri Stellar Intensity Interferometer. Besides, active mirror control allows to split the primary mirrors into sub-mirrors. This allows to make simultaneous calibration measurements of the zero-baseline correlation or to simultaneously collect six baselines below 17 m with almost arbitrary orientation, corresponding to angular scales of $\sim$1-50 mas. We plan to perform test observations adding the nearby Cherenkov Telescope Array (CTA) LST-1 23~m diameter telescope by next year. All three telescope pairs will be correlated simultaneously. Adding LST-1 is expected to increase the sensitivity by at least 1~mag and significantly improve the u-v plane coverage. If successful, the proposed correlator setup is scalable enough to be implemented to the full CTA arrays.
\end{abstract}


\section{INTRODUCTION}
\label{sec:introduction}  

After Hanbury-Brown \& Twiss\cite{HBT1956} successfully proved that at sufficiently short baselines, the spatial correlation measurements of stars exhibit a photon bunching signal for measurements within a defined area on the ground, and that this area depends on the angular size of the star, an interferometer in Narrabri, Australia \cite{HBbook1974} was used to measure the diameters of the 32 brightest stars in the Southern Hemisphere. This ''intensity interferometry'' technique was however brought to a stop because progress called for more light collection or faster photodetectors and electronics, hard to achieve at the time.

The situation has changed in the last decade. Faster Avalanche Photo Diodes have been recently applied to intensity interferometry (see \cite{Guerin2018} and these proceedings) and already running Imaging Atmospheric Cherenkov telescopes (IACTs) are equipped with large reflectors ($>$100~m$^2$) and fast acquisition chains ($\sim$1~ns). What is more, the new generation IACT observatory, called Cherenkov Telescope Array (CTA), is under construction (see references \cite{Dravins, VERITAS, magic_2019, VERITAS2, cta}).

\section{CURRENT INSTRUMENTAL SETUP}
\label{sec:instrumental_setup}  

MAGIC is a system of two IACTs located at the Roque de los Muchachos Observatory on the island of La Palma in Spain\cite{upgrade1}. Equipped with 17~m diameter parabolic reflectors and fast photomultiplier (PMT) cameras, the telescopes record images of extensive air showers in stereoscopic mode, enabling the observation of very-high-energy $\gamma$-ray (VHE) sources at energies of few tens of GeV \cite{upgrade2}.

In April 2019, to prove that MAGIC was technically ready to perform intensity interferometry observations, a test was performed using MAGIC telescopes and an oscilloscope as a readout\cite{magic_2019}. Using MAGIC standard photo-detectors and camera electronics, temporal correlation was detected for three different stars of known angular diameter. The sensitivity and the degree of correlation were consistent with the stellar diameters and the instrumental parameters. The acquisition however was subject to a low duty cycle and the mechanical installation of the optical filters did not allow for a quick change of setup between VHE and interferometry observations.

In the following we describe the technical modifications that have been implemented in the telescopes to enable them for intensity interferometry. Some of these modifications have been partly introduced elsewhere\cite{spie2020, icrc2021}. A key consideration was always not to affect regular VHE observations and to allow to switch from ''VHE observation mode'' to ''interferometry observation mode'' and back in less than one minute.

\subsection{Mirror and active mirror control}
\label{subsec:mirror}

The MAGIC reflector follows a parabolic shape to minimize the time spread at the focal plane. The focal length is 17~m. The reflector is formed by $\sim$250 1~m$^2$ spherical mirror tiles. Because the mirrors are supported by a light-weight space-frame of carbon fiber and aluminum tubes, the reflector shape deforms with elevation and each mirror tile is equipped with two actuators to compensate for this deformation. This so-called Active Mirror Control (AMC) adjusts the mirror in a few seconds and runs typically every 20 minutes. Approximately 70\% of the light of a point source is focused on a pixel.

We strongly profit from the AMC during intensity interferometry observations. On one hand, the reflectors are not focused to stars at infinity during VHE observations but we can change focus to infinity in a matter of seconds before switching to interferometry.

\begin{figure} [ht]
  \begin{center}
    \begin{tabular}{c c} 
      \includegraphics[height=6cm]{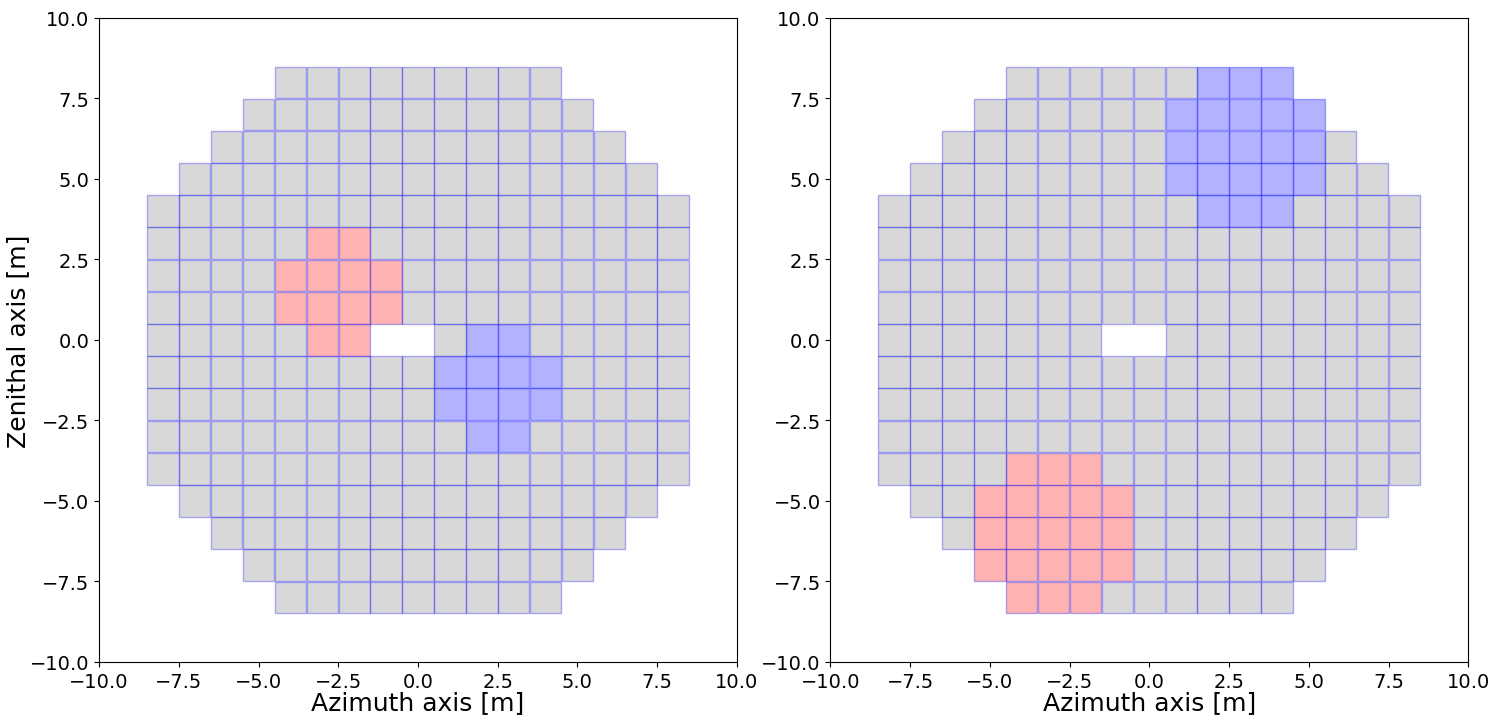}
    \end{tabular}
  \end{center}
  \caption[submirror-examples]
          { \label{fig:submirror-examples}
            Two examples of submirror configurations. Both panels show a sketch of the 1~m$^2$ mirror tiles in one of the MAGIC telescopes. As we shall see, we use two camera pixels during interferometry observations. The tiles colored in red are focused on one of them while the tiles colored in blue are focused to the second one. The remaining tiles are focused to the center of the camera, i.e. not employed in the observation. The two submirrors have an area of 12~m$^2$ in the left panel and 21~m$^2$ in the right panel.}
\end{figure}
On the other one, we have modified the AMC software so that it can focus the entire reflector or arbitrary groups of mirror tiles to arbitrary positions in the camera focal plane (within a circle of roughly 1 deg radius). We can define configuration files defining which panels are focused to which pixel in the camera.

\begin{itemize}
\item In the simplest configuration the entire reflector is focused on a pixel that is 0.8 deg from the camera center.
\item  In the so-called chessboard configuration half of the mirror tiles (arranged like the black squares in the chessboard) are focused to a pixel while the other half (the white squares) are focused on a second pixel. This allows to study the correlation between the white and black mirrors, which corresponds to a distribution of baselines always $<$17~m, providing an almost direct way of measuring the zero-baseline correlation.
\item  In other configurations, two roundish groups of 12-21 tiles are focused to each of the two pixels: see for instance the two configurations in figure \ref{fig:submirror-examples}. This was inspired by the I3T concept\cite{I3T} and allows to measure the correlation for a specific baseline $<$17~m and different orientations in the sky.
\end{itemize}

\subsection{Optical filters and filter holders}
\label{subsec:filter_holder}

The signal to noise of the correlation of the telescope signals is insensitive to the width of the optical passband of the detected light. However a filter spectral response with sharp spectral cutoffs improves the sensitivity of the measurement. Even more importantly, we are observing bright stars that, accounting for MAGIC photo-detection efficiency, would damage the PMTs after a short exposure time. Therefore interferometry observations require installing filters in front of the pixels connected to the correlator. We are using interference filters manufactured by Semrock of model 425-26nm.

\begin{figure} [ht]
  \begin{center}
    \begin{tabular}{c} 
      \includegraphics[height=7cm]{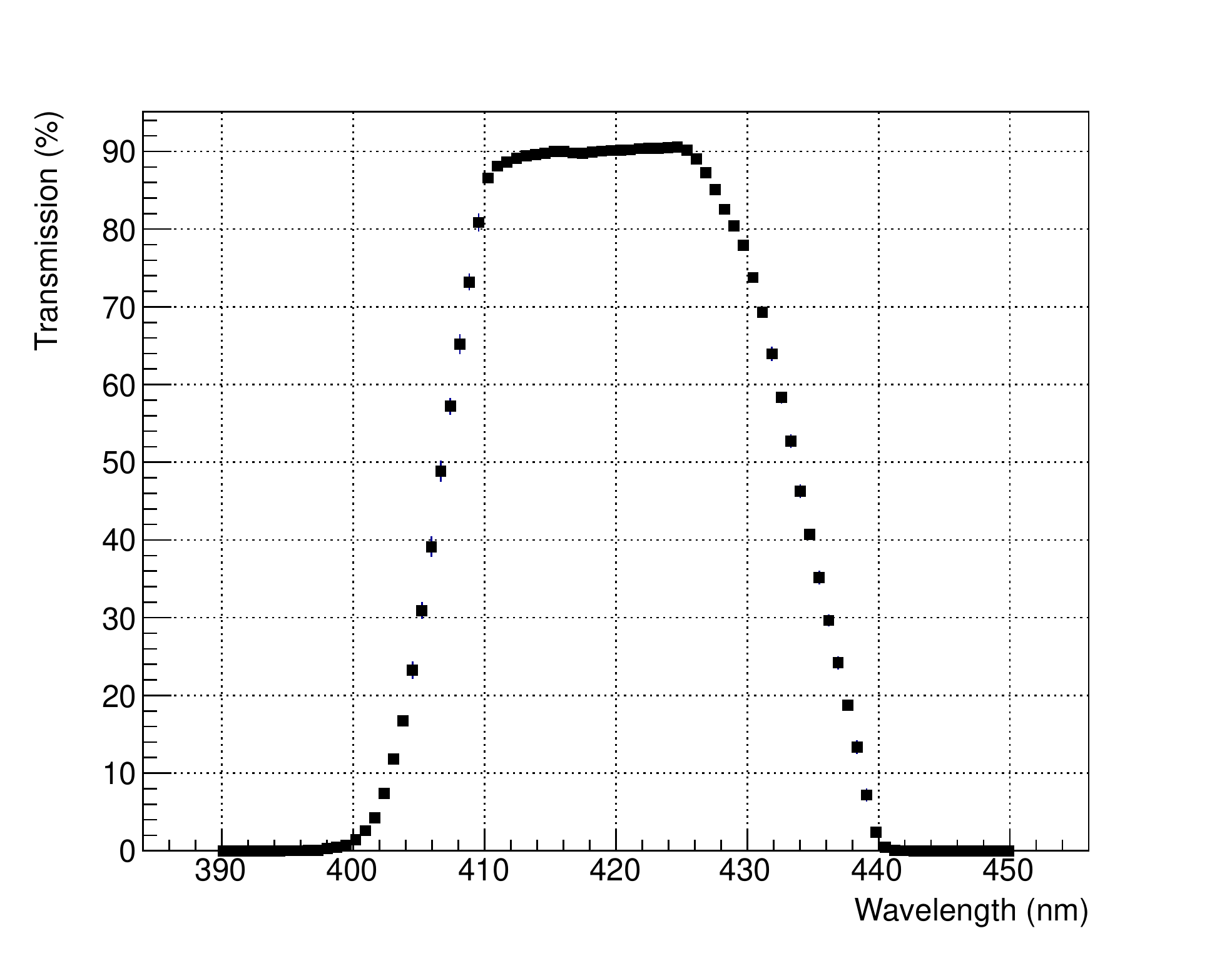}
    \end{tabular}
  \end{center}
  \caption[transmission-curve]
          { \label{fig:transmission-curve}
            Transmission curve of the Semrock 425-26nm optical filter for light collected by the MAGIC reflector.}
\end{figure}

The spectral transmission curve for incident parallel light is centered at 425 nm and has a FWHM of 25~nm with relatively sharp edges. This shape is strongly modified in our setup because MAGIC has a small f/D. We have used the MyLight modeling online tool provided by Semrock to calculate the effective spectral transmission curve in a cone of half angle 26.5$^{\circ}$. This curve is shown in figure \ref{fig:transmission-curve}.

\begin{figure} [ht]
  \begin{center}
    \begin{tabular}{c} 
      \includegraphics[height=9cm]{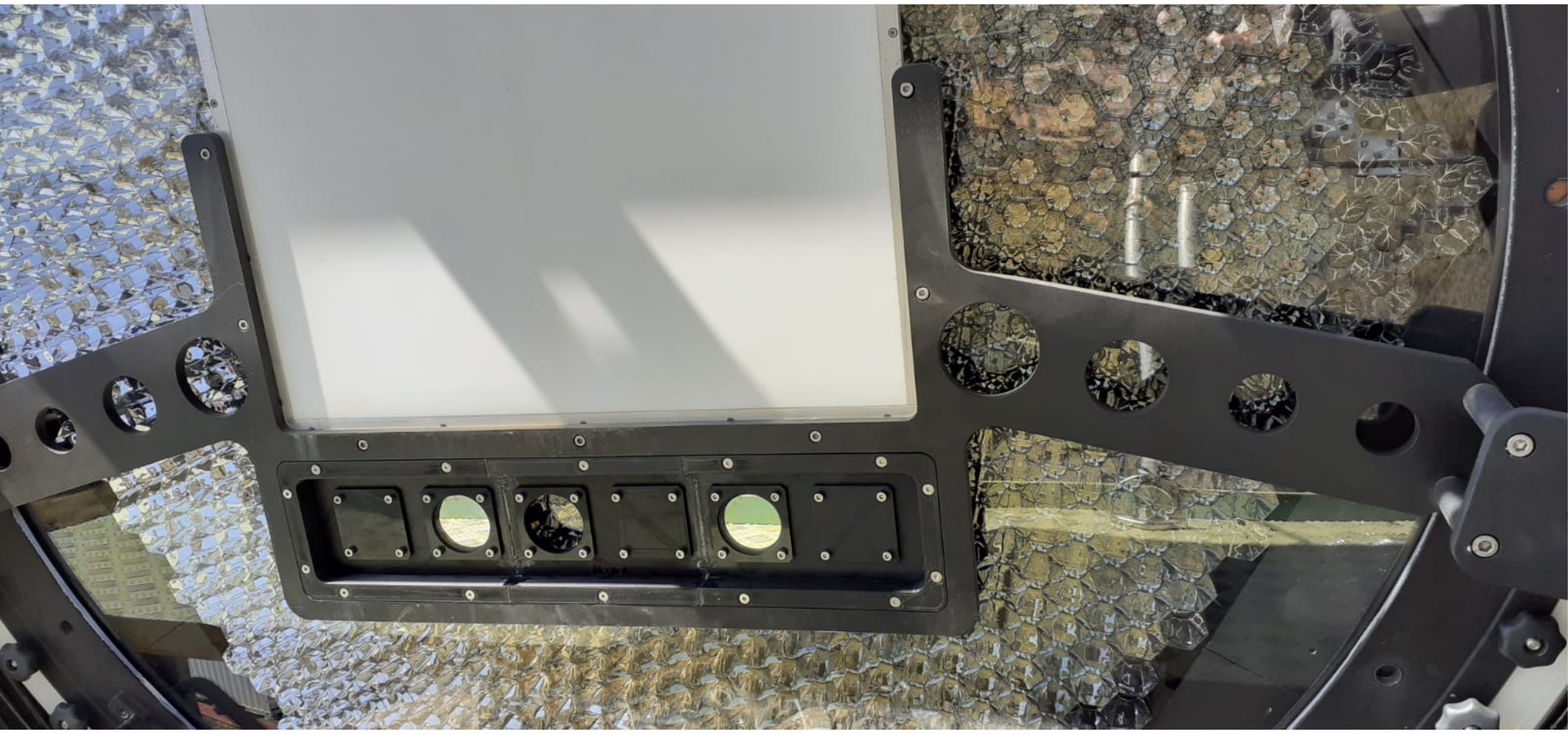}
    \end{tabular}
  \end{center}
  \caption[filter-holder]
          { \label{fig:filter-holder}
            Filter holder of the MAGIC telescopes. The rectangular part at the bottom of the white surface is the mechanical structure holding the filters in front of the camera PMTs. They have room for six filters. At the time the picture was taken only two filters were installed (greenish circles), one of the holes was left open and three holes were closed with a plastic cap. Behind the filter holder and the diffusive white target one can see the hexagonal light concentrators right in front of the PMTs.}
\end{figure}

The filters have a diameter of 50~mm and are held 20~mm in front of the pixel using a mechanical frame (''filter holder''). This frame piggybacks on an existing mechanical structure that holds a diffusive plate used for telescope calibration purposes (the so-called ''white target''). This structure is operated remotely and can be deployed in front of the PMTs in a matter of seconds. Figure \ref{fig:filter-holder} shows a picture of the filter holder under the white target. There is room for six filters in a horizontal line, allowing simultaneous signal and background monitoring using identical filters during interferometry observations. Each filter is placed and centered in front of a PMT.

\subsection{Light detection and signal transmission}

Each telescope is equipped with a 1039-pixel PMT camera at the primary focus.
The PMTs are 25.4~mm diameter and have 6 dynodes. The relatively small number of dynodes
has been selected in order to deliver a low amplification gain and to achieve shorter pulses. A low gain avoids ageing when
the telescopes take VHE data in the presence of strong Moonlight background and is
also convenient for interferometry observations of bright stars.
A hexagonal shape Winston Cone is mounted on top on each PMT. The distance between PMT centers is 30~mm and corresponds to a 0.1$^{\circ}$ FOV.

The electrical signals of the PMTs are amplified (AC coupled, $\sim$25 dB amplification) and then transmitted via independent optical fibers by means of vertical cavity surface emitting lasers (VCSELs) to a separate readout location. The fibers are $\sim$162~m long. The 850~nm multimode optical signal is converted to an electrical signal at the readout end of the fiber.

Only two pixels (numbers 251 and 260) in each of the two cameras are currently connected to the correlator. The optical fibers connected to these pixels have been disconnected from the VHE data acquisition. Each of the four pixel fibers is connected to fiber delays of different length terminated with a GaAs photodiode from IMM Photonics in an individual electromagnetically shielded enclosure. The photodetector signal is amplified by a Femto HSA-Y-2-40 amplifier with 40dB gain and 2 GHz bandwidth. The signal is then digitized and correlated (see next section).

The PMT pulses have a width of 2.5~ns. The front-end electronics and optical transmission have a bandwidth of $>$300 MHz. The bandwidth of the acquisition chain is limited by the digitizer to $\sim$130 MHz.

\subsection{Digitizer and correlator}

The correlator hardware and software have been designed to harness the massively parallel nature of state-of-the-art GPUs to process in real time the data captured using two fast digitizer boards Spectrum M4i.4450-x8 PCIe 2.0.

Each digitizer deals with two channels providing up to 500 MS/s of simultaneous sampling rate with a resolution of 14 bits per sample. In addition, the chosen digitizers support remote-direct-memory access (RDMA), that allows direct data transfers between digitizer’s memory and GPU’s, avoiding intermediate copies and, therefore, increasing throughput and lowering latency. The two Spectrum boards clocks are synchronized by means of the Star-hub module attached to the carrier card in the correlator chassis.

The correlator is implemented in a computing server with off-the-shelf hardware: two processors (20 cores in total), SSDs for fast access and HDDs for longer term storage and tests, and a Nvidia Tesla V100 GPU. The GPU chosen is the PCIe 3.0 x16 model with 5120 cores, 32 GB HBM2 RAM and 14 TFLOPs of single-precision performance.

The correlator software computes the correlation as a function of the delay between pairs of channel in a wide delays window by means of the convolution theorem in Fourier space using the Fast-Fourier-Transform (FFT). It is developed in CUDA C using the Nvidia FFT library for the convolution computation, and the Spectrum CUDA Access for Parallel Processing (SCAPP) SDK for transfering the data between the digitizers and the GPU in streaming mode. We are using double precision for the calculations.

The correlator software is divided in three parts:
\begin{enumerate}
\item Firstly {\bf the initialisation section} configures several parameters of the digitizers and GPU  (number of channels, acquisition rates, input paths and ranges, RDMA, execution time, etc.), initialises the data structure in both the CPU and GPU, and creates the thread that writes the resulting correlation to a storage media. Then the data processing loop is started.

\item {\bf The data processing loop} runs continuously for the programmed duration of data taking, or until an error occurs or it is interrupted by the operator. Within this loop the code first checks and retrieves  a frame of a given size of incoming data from the digitizers. Then the retrieved samples are reordered within an input buffer  and prepared for the convolution computation buffer according to channel. Simultaneously, some per-channel statistics are obtained (means and standard deviations), and finally the corresponding convolution is computed, normalized and added to an output buffer. The input and output are implemented using double buffering, which allows to keep a live-time of the correlator of 100\%.

\item Every second the output buffer is swapped, and a {\bf writer thread} is awaken that saves the convolution results and any obtained statistics to disk and goes back to sleep.
\end{enumerate}

The correlation code has been used to process data from the two digitizers in real time at 4 GB/s. Alternatively the code can dump raw data from one digitizer (two channels) to disk in real time for tests.

\section{FIRST OBSERVATIONS AND ANALYSIS RESULTS}
\label{sec:observations_results}  

The full instrumental setup described in the previous section was only fully realized in July 2021. Unfortunately a volcano eruption near ORM prevented us from taking sky data until January 2022. In the last months we have collected more than 250 hours of observations taken during Full Moon time. On one hand, we have observed bright stars (typically B$<$2$^m$) whose diameter is known from other interferometric measurements. We will use these stars to calibrate the instrument and understand its performance. On the other one, we have observed around 10 stars of unknown diameter.

Our readout is able to correlate signals from four acquisition channels. The four channels are currently connected to pixels 251 and 260 of the two telescopes. We refer to these channels as A, B for one of the telescopes, and C and D for the second one. As described in subsection \ref{subsec:mirror} we are taking data in three different observation modes: focusing the full mirror to one pixel (''full mirror observations'', either with pixel 251 or 260), focusing half of the mirror to pixel 251 and the other half to pixel 260 (''chessboard observations''), and focusing a sub-mirror to 251 and a second one to 260.

We refer the reader to reference \cite{magic_2019} where we explained in detail how we extract the correlation signal and determine the star's visibility curve. Here we will only sketch the generalities of the method.

Since the PMT signals are AC-coupled we cannot directly calculate the intensity correlation but we must calculate Pearson's correlation factor $\rho$ and later divide by the PMT anode currents produced by the star (DC$_1$ and DC$_2$) to account for changes in star flux or signal detection efficiency. This observable is proportional to the squared visibility V$^2$:
\begin{equation}
  V^2 = K \frac{\rho}{\sqrt{DC_1 DC_2} }
  \label{eq:V2_and_rho}
\end{equation}
where $K$ is constant if the gain of the PMTs (i.e. the high voltage) remains constant. As DC$_1$ and DC$_2$ need to account only for the light of the star, we need to subtract the DC associated to the night-sky background (NSB). As these observations are generally performed during bright-Moon periods (in which standard VHE observations are not performed) the NSB contribution may be significant, specially for fainter stars. The holder described in section \ref{subsec:filter_holder} allows to devote PMTs, behind identical filters as those used for computing the correlation, to simultaneously estimate the DC associated to the NSB.

\begin{figure} [ht]
  \begin{center}
    \begin{tabular}{c} 
      \includegraphics[height=9cm]{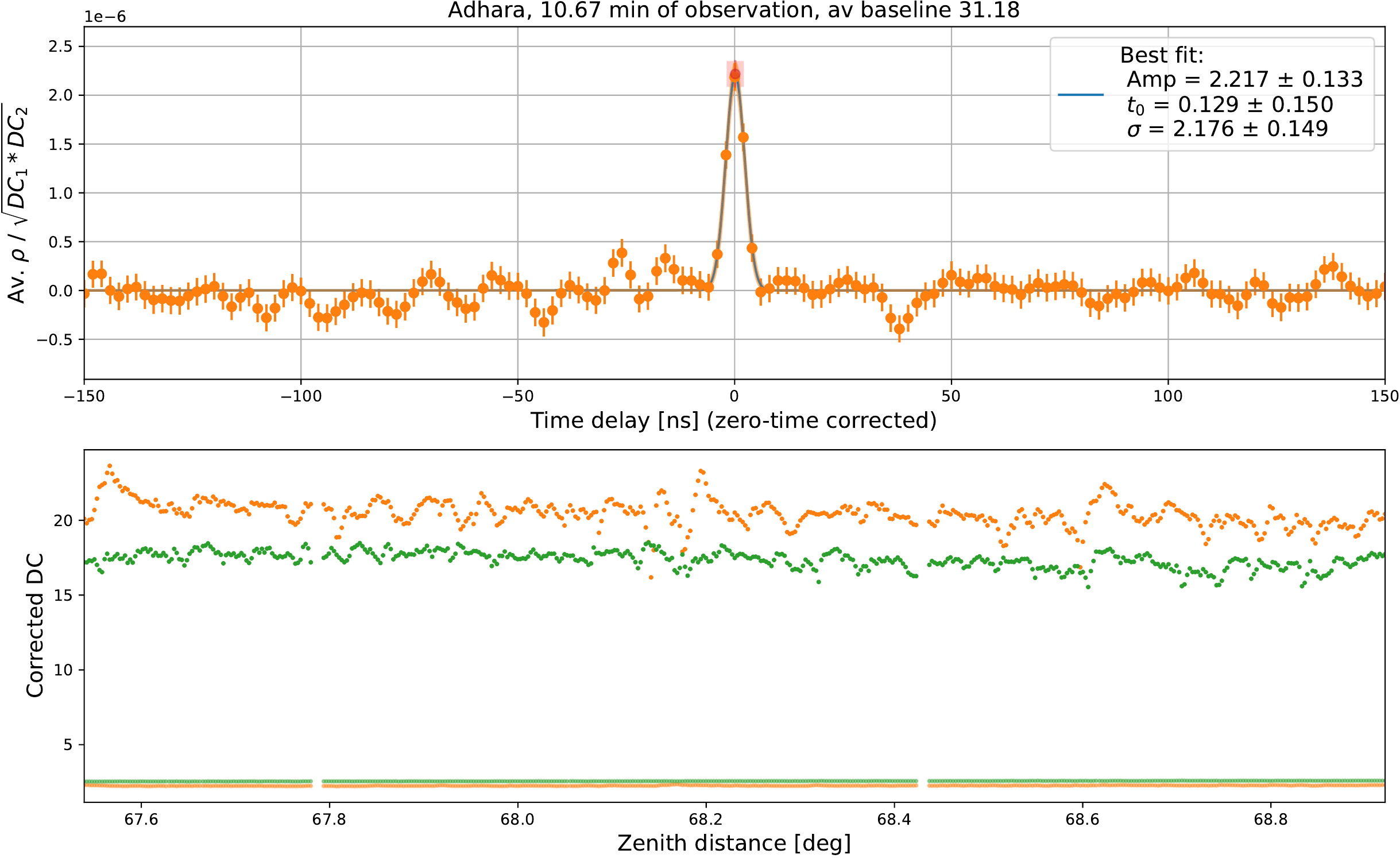}
    \end{tabular}
  \end{center}
  \caption[example-correlation-peak-DCs]
          { \label{fig:example-correlation-peak-DCs}
            The upper panel shows Pearson's correlation factor as a function of time delay between the PMT signals during an observation of the star Adhara corrected with the DC currents of the pixels used. Delay zero corresponds to the expected delay between the pixels for the position of the star in the sky. The lower panel shows the evolution of the PMT anode currents (DCs) of both the star (points above 15 $\mu$A) and background during the observation as a function of the zenith distance. The fluctuations of the star DC can be explained by lags in the drive control loop and wind gusts.
          }
\end{figure}

Pearson's correlation, $\rho$, is actually calculated simultaneously for a large range of time delays. The correlation signal is observed only for the time delay corresponding to the telescope orientations while the rest of the delays are used to estimate the expected background coming from spurious electronic correlations.
Figure \ref{fig:example-correlation-peak-DCs} shows an example of the correlation signal detected over $\sim$ 10 min of observation time for one of our calibration stars (Adhara).
We fit $\rho(\tau)$ to a gaussian and use the gaussian amplitude to estimate the correlation factor for that specific observation. The uncertainty is estimated from the fluctuations of $\rho$ in a range of time delays where the signal is not expected.

\begin{figure} [ht]
  \begin{center}
    \begin{tabular}{c} 
      \includegraphics[height=8cm]{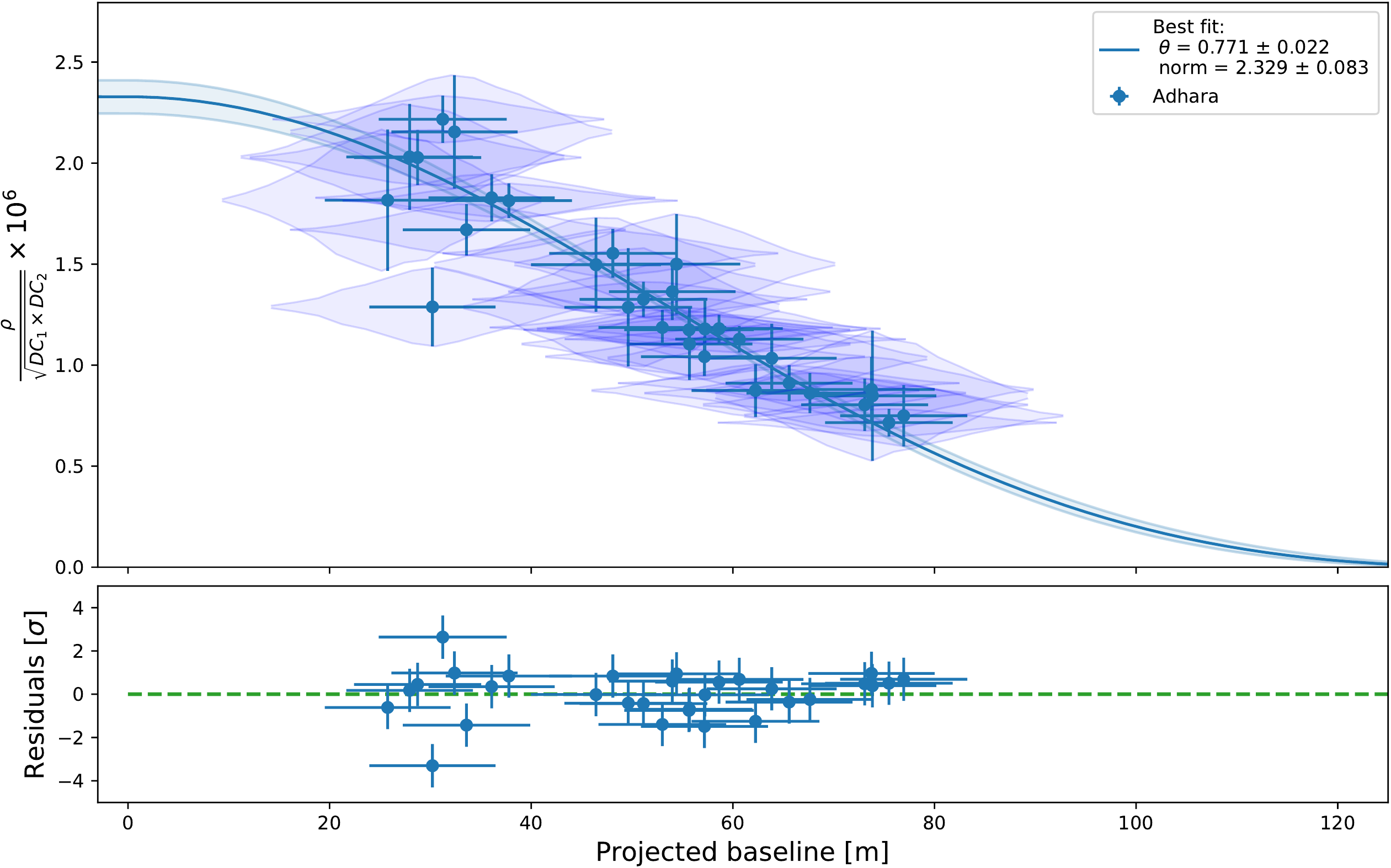}
    \end{tabular}
  \end{center}
  \caption[example-visibility-fit]
          { \label{fig:example-visibility-fit}
            The upper panel shows Pearson's correlation factor corrected by the PMT DCs (in $\mu$Ampere) as a function of baseline during a set of observations of the star Adhara. Data is divided in uniform steps of 2.5 m in baseline. The shaded areas shown for each point are defined in vertical by the statistical error of the normalized Pearson's correlation amplitude and in horizontal by the baseline distribution of each observation, calculated from the mirror tiles used and their relative distance. X axis error bars are the 68\% containment radius of the baseline distribution of each point. The points have been fit to the visibility function of an uniform disk profile, where each point is weighted according to its normalized baseline distribution. The results of the fit are displayed. The lower panel shows the residuals to this fit, in units of their standard deviation (where, by definition, all Y-axis error bars are 1).
          }
\end{figure}
We plot $\frac{\rho}{\sqrt{DC_1 DC_2}}$ as a function of telescope baseline and fit it to the visibility curve expected from a uniform disk to determine the diameter of the star $\theta$. Figure \ref{fig:example-visibility-fit} illustrates this fit for the same star Adhara and a set of observations.

The measured visibility depends on the range of baselines covered by the observation, which is affected not only by source tracking, but also by the size of our mirrors. We calculate the distribution of baselines associated to each observation and, during the model fitting procedure, we compare the measurements with the baseline-weighted average of the model. This effect is significant for short baselines and was not considered in our previous publications.

A key parameter in the visibility fit is the correlation at zero baseline or ''zero-baseline correlation'' (ZBC). This parameter depends on the telescope and pixel parameters but should be stable in time. It can be extracted from the visibility fit to calibration stars and then introduced as a constant in the fits to target stars of unknown diameters.

\begin{table}[ht]
  \label{tab:ZBC-all-pairs}
  \begin{center}
    \begin{tabular}{|l|l|l|}
      \hline
      \rule[-1ex]{0pt}{3.5ex}  Pixel pair & Diameter & ZBC  \\
      \hline
      \rule[-1ex]{0pt}{3.5ex}  AC & 0.77$\pm$0.03 & 2.29$\pm$0.11   \\
      \hline
      \rule[-1ex]{0pt}{3.5ex}  AD & 0.80$\pm$0.06 & 2.2$\pm$0.2  \\
      \hline
      \rule[-1ex]{0pt}{3.5ex}  BC & 0.80$\pm$0.07 & 2.4$\pm$0.2 \\
      \hline
      \rule[-1ex]{0pt}{3.5ex}  BD & 0.76$\pm$0.04 & 2.39$\pm$0.13  \\
      \hline
    \end{tabular}
  \end{center}
  \caption{Uniform disk diameter and ZBC extracted from the visibility fits of the star Adhara calculated from the four different correlation pairs used during chessboard mode. Diameter is in milliarcsec and ZBC is in units of $\mu$A$^{-1}$. The diameter of Adhara has been reported\cite{HB1974} to be 0.77$\pm$0.05 mas. Note that the uncertainties are not the same for all pairs because we are adding together runs in chessboard mode (where all pairs have a signal), full mirror to 251 and full mirror to 260 (where only the corresponding pair has a signal) so each pixel pair can correspond to a different observation time and different baseline distribution.}
\end{table}
Alternatively we may take data on target stars with four acquisition channels in chessboard mode. In this configuration we measure the correlation between six different channel pairs. The channel pairs in different telescopes (AC, AD, BC, BD) correspond to one baseline $>$17~m which depends on the telescope orientation and changes along the night, whereas the channel pairs in the same telescope (AB and CD) correspond to a distribution of baselines $<$17~m which remains constant and is expected to be a good estimate of the ZBC. Hence the chessboard configuration would allow to make a simultaneous measurement of the visibility at zero baseline and at a baseline of a few tens and meters, which would allow by itself to estimate the star diameter without a calibrator star.

Table 1 shows the ZBC and diameter extracted from visibility fits data taken from the calibrator star Adhara using all four different telescope channel pairs. Both the reconstructed diameter and ZBC values are consistent for all channel pairs. This gives us confidence that the signal extraction procedure and instrument setup are robust.

Unfortunately the correlation of channels in the same telescopes (AB and CD) suffer from a strong correlated electronics noise. The noise has a clear structure in time and seems to be stable, so preliminary attempts to remove it via software filtering are encouraging. This is still work in progress, so at the moment we cannot provide a reliable measurement of ZBC using same telescope correlations.

\begin{figure} [ht]
  \begin{center}
    \begin{tabular}{c} 
      \includegraphics[height=8cm]{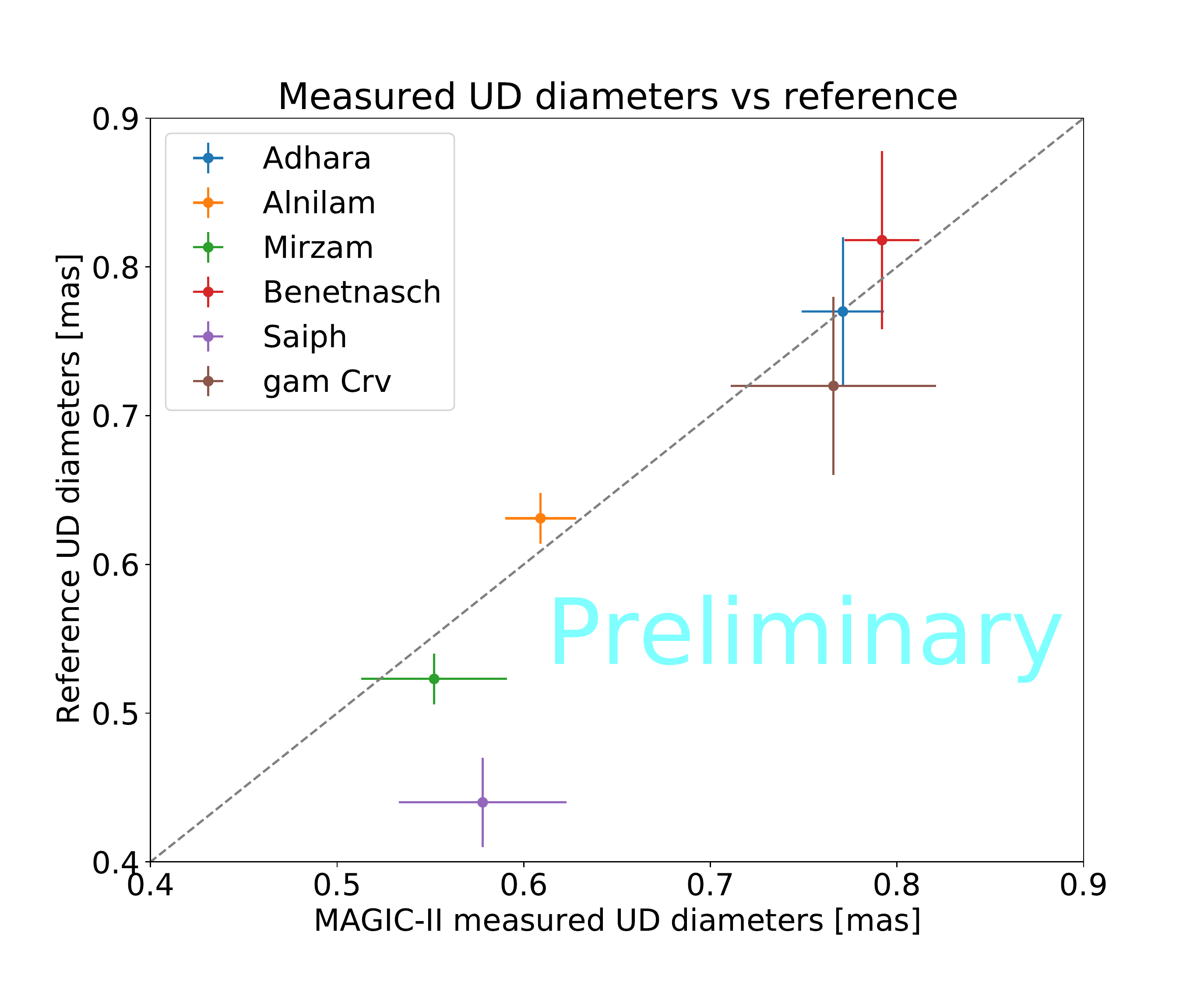}
    \end{tabular}
  \end{center}
  \caption[diameter-calibrator-stars]
          { \label{fig:diameter-calibrator-stars}
            Uniform disk diameter measured with other interferometers as a function of diameter measured with our interferometer for several calibrator stars. The reference diameters of Adhara, Saiph and $\gamma$~Crv are from ref\cite{HB1974}, Alnilam and Mirzam from ref\cite{VERITAS} and Benetnasch from ref.\cite{NPOI}
          }
\end{figure}
Figure \ref{fig:diameter-calibrator-stars} compares the diameters measured with our instrument to the diameters measured with other interferometers for six calibrator stars. All diameters are consistent with previous measurements except for the star Saiph. These measurements are still preliminary. We are currently estimating the scale of our systematic uncertainties.

Once we are confident with the diameter extraction method using calibrator stars we will proceed to process the data of the sample of stars of unknown diameter. We are able to advance that most of the stars we observed show a clear correlation signal and the weakest star observed so far has a B magnitude close to 4, showing a significant detection.
In fact, from these measurements we can tentatively conclude that the significance of the correlation signals is consistent with the expectations that are discussed in the last section of this paper and displayed in figure \ref{fig:error-diameter-MAGIC-4LST}.

We have also made exploratory measurements of stars with a diameter larger than 1~mas. For such stars the distance between the telescopes is beyond the first zero of the visibility curve but we can still measure the correlation using submirrors inside each telescope. Unfortunately the area of the submirror is smaller than the area of the full reflector and the SNR of the signal decreases with the area. What is more, we suffer from the above-mentioned electronic noise in same telescope data. However we observe clear correlation signals for Sirius and submirrors of 12~m$^2$ and 24~m$^2$. We will show the corresponding preliminary visibility curves at the conference.

\section{UPGRADE PLANS}
\label{sec:upgrade}  

On considering upgrades to the current setup it is useful to recollect equation 5.17 in reference \cite{HBbook1974}~, which allows to calculate the significance (signal over noise) of the correlation for a pair of identical telescopes, a given experimental setup and unpolarized light. The equation can be written as:
\begin{equation}
  S/N = \\
  A \cdot \alpha(\lambda_0) \cdot q(\lambda_0) \cdot n(\lambda_0) \cdot |V|^2(\lambda_0,d) \cdot \sqrt{b_v}
  \cdot F^{-1} \cdot \sqrt{T/2} \cdot \sigma
  \label{eq:significance}
\end{equation}
where A is mirror area, $\alpha(\lambda_0)$ is the quantum efficiency (QE) of the PMTs for the filter's central wavelength $\lambda_0$, q$(\lambda_0)$ is the QE of the remaining optics, $n(\lambda_0)$ is the star's differential photon flux, $b_v$ is the effective cross-correlation electrical bandwidth, $F$ is the excess noise factor of the PMT and $T$ is the observation time.
Finally $\sigma$ is the normalized spectral distribution of the light after the filter as defined in formula (5.6) of ref\cite{HBbook1974}~. $\sigma$ would be equal to 1 if the filter transmission curve is a boxcar function and the spectrum of the light is flat whereas it reduces as the curve rising and falling edges of the curve gets broader (it is in fact 0.86 for the transmission curve in figure \ref{fig:transmission-curve}). We have assumed that both telescopes are identical and we have neglected the effect of the Moon and additional noise in the readout chain.

When dealing with an array of N$_{tel}$ such telescopes the S/N increases with the square root of the telescope pairs. The number of telescope pairs is $N_{tel}(N_{tel}-1)/2$ so S/N increases roughly linearly with N$_{tel}$.

On the other hand equation 5.17 is remarkable in the fact that the S/N does not depend on the ''optical passband'' (the width of the spectral range). This means that one could split the light into many spectral channels N$_{spectral}$ of a much smaller passband and each of these channels would produce a signal with the same S/N. By combining all the spectral channels the total S/N of the interferometer increases by a factor $\sqrt{N_{spectral}}$.

Hence one can improve the sensitivity of the interferometer by improving a number of parameters: the number of telescopes, their mirror area, the QE of the photodetectors, the bandwidth of photodetector and readout, and the number of spectral channels. In what follows we will review some of our plans to increase some of these parameters.

\subsection{A scalable correlator}

When increasing N$_{tel}$ or N$_{spectral}$ the correlator must be dimensioned accordingly. The computation power only increases linearly with N$_{spectral}$. The real challenge is that the computation power increases quadratically with N$_{tel}$ because one needs to calculate the correlation of each individual telescope pair. Here we are describing a correlator architecture that is able to scale up to a large N$_{tel}$ because computation power increases essentially linearly with it. We are building on ideas developed for radio interferometry (see e.g. this review\cite{radio2020}).

\begin{figure} [ht]
  \begin{center}
    \begin{tabular}{c} 
      \includegraphics[height=7cm]{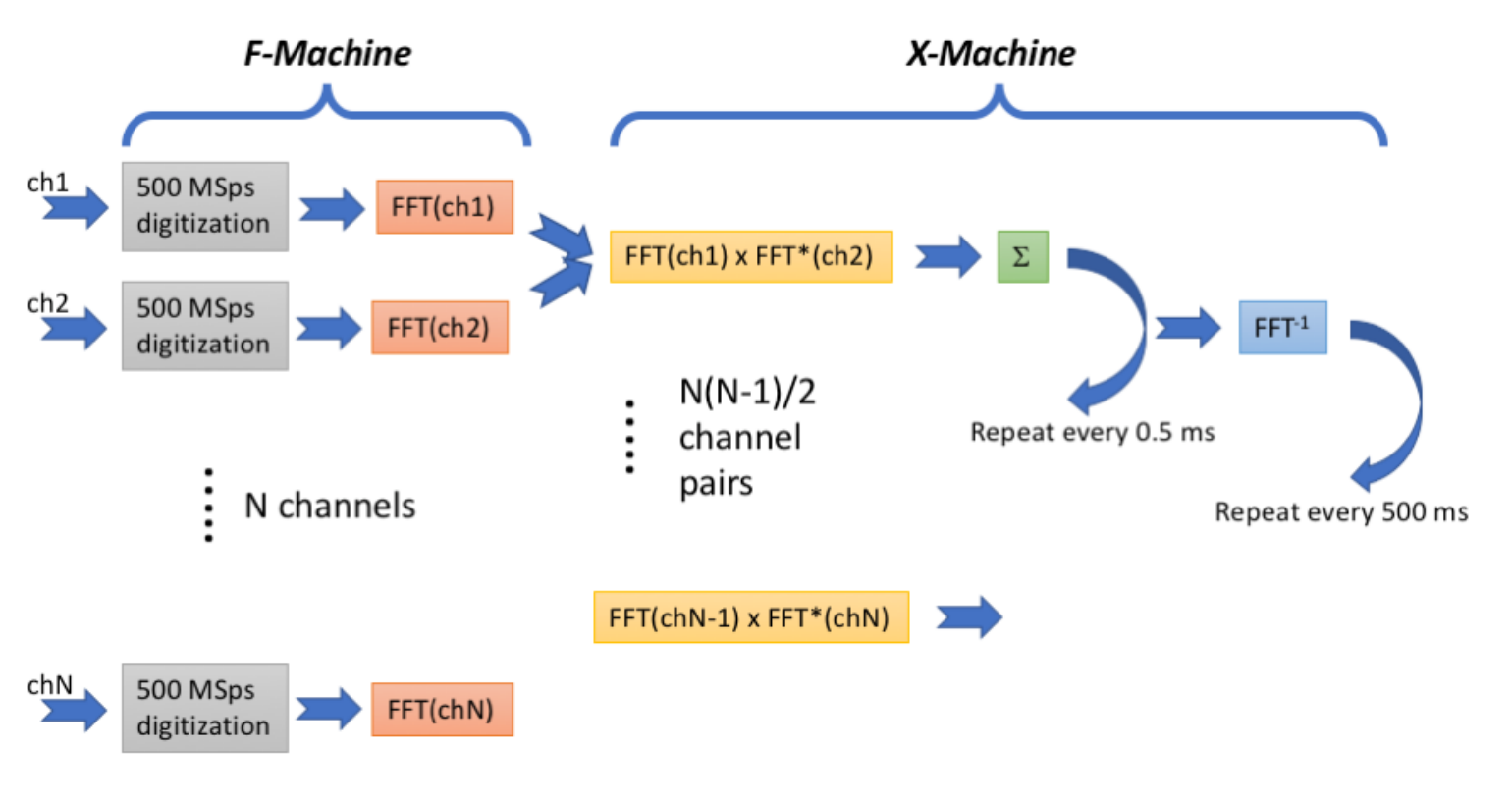}
    \end{tabular}
  \end{center}
  \caption[scalable-correlator-concept]
          { \label{fig:scalable-correlator-concept}
            Architecture of a correlator with N independent channels.
          }
\end{figure}

Figure \ref{fig:scalable-correlator-concept} describes how the correlations between all possible 2-channel pairs are performed. In a first step the signal from each camera pixel is digitized. There are N of these channels. A Fast Fourier Transform is applied over a defined time interval (a ''frame'', of 500 $\mu$s duration) on each channel. Finally, the FFT is encoded as a Hartley transform to reduce the amount of information buffered. We refer to these first two steps as an ''F-Machine''.

The N encoded FFTs are then sent to the ''X-Machine'' where they are decoded. The number of channel pairs is N(N-1)/2. For each pair of channels i and j, we multiply the FFT(ch i) and the conjugate FFT*(ch j). The result is summed to a buffer that accumulates the summation of all previous frames. The process repeats for 1000 frames (i.e. 500 ms) and then an inverse FFT is applied to the sum. From the inverse FFT we can easily calculate an array of correlation coefficients vs time delay of the signal in the channel pair. Only a small portion of the array is saved to disk.

Let us see how computation power scales with N. The F-Machine only calculates FFTs for N channels so the computation power scales with N. Instead the number of calculations in the X-Machine scales with N(N-1)/2. However the operation FFT*FFT$^*$ demands very little power compared to the inverse FFT, and, since the inverse FFT is applied over the sum of a large number of frames, the computation power of the X-Machine is effectively reduced by the number of frames it accumulates. The net result is that the F-Machine drives the requirement for computation power and the latter scales linearly with N. In first approximation this holds for as long as N is smaller than the number of frames, in our case 1000.

The concept of calculating the inverse FFT over the average of a large number of frames has been tested successfully in our current interferometer, and is routinely used during our standard observations.

We are considering a specific hardware implementation with the following elements:
\begin{itemize}
\item Same digitizers that we are using right now in MAGIC: PCIe 2.0 card with 500 MSps (Spectrum M4i.4450-x8), 14 bit resolution and two channels.
\item We foresee one computer for each Spectrum card, i.e. 2 channels per computer and 10 computers for a total of 20 channels.
\item Each of these computers (F-Machine computers) will also be equipped with a GPU to perform the two FFTs.
\item One single server for the X-Machine, equipped with two GPUs: the first one calculates multiplications and sums, while the second one calculates the FFT$^{-1}$.
\item A data throughput of 4 GByte/s (32 Gbit/s) must be possible from the F-Machine computer to the X-Machine server.  We plan to use RDMA and GPUDirect to transfer data directly from GPU to GPU. This can be implemented with technology of the company Mellanox.
\item An F-Machine GPU needs to execute 800 GFLOPs (2 FFTs) while the first X-Machine GPU need 280 GFLOPs (190 multiplications and sums, the final FFT$^{-1}$ adds hardly any computation load). Even if these demands can be met with a low-end GPU, we would rather go a scientific GPU that offers between 10 and 100 times more computing power than the demands, both in order to stay on the safe side and to ensure that we can run an offline data analysis when we are not taking data, specifically for calculation crosschecks and to extract images out of the correlation arrays.
\end{itemize}

We have recently purchased the elements to build a test correlator with only four channels. The F-Machine is made of two computers each with a 2-channel digitizer, and a computer for the X-Machine. All three computers are Supermicro AS-2014CS-TR (AMD ROME 7352 DP/UP 24C/48T processor, SSDs, PCIe4 bus). We have selected GPUs Nvidia Ampere A100.
The three computers are linked with a Mellanox MQM8790-HS2F 40-port Infiniband switch (200Gbit/s maximum bidirectional data rate), a Mellanox card MCX653106A-HDAT-SP in each computer and optical fibers. We are currently setting up the correlator for a lab performance test.

If the test correlator performs successfully more computers can later be added to the F-Machine. The digitizer may in fact be upgraded to a new version of the same company with 1 GSPs sampling. We are actually planning to apply this correlator concept to the CTA LSTs that we are discussing below.

\subsection{Extending the interferometer to the LSTs}

CTA\cite{cta} will consist of two arrays of IACTs at the northern and southern hemispheres. With its large number of telescopes and different baselines, equipping CTA for intensity interferometry will deliver a significant increase in performance respect to the current IACT arrays (see for instance section 14.3 of reference \cite{cta} for some of its scientific goals). However there is no well defined timeline or specific technical solution for the CTA interferometer.

\begin{figure} [ht]
  \begin{center}
    \begin{tabular}{c} 
      \includegraphics[height=8cm]{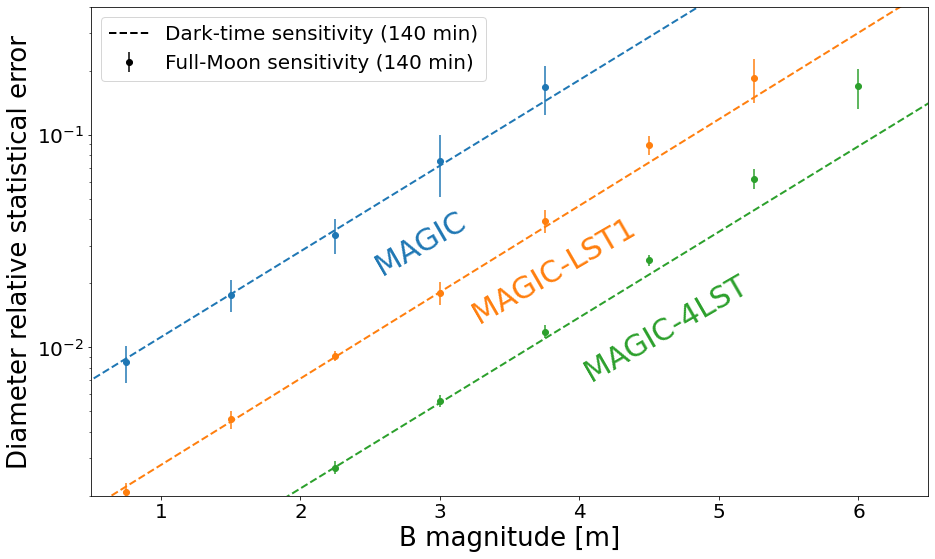}
    \end{tabular}
  \end{center}
  \caption[error-diameter-MAGIC-4LST]
          { \label{fig:error-diameter-MAGIC-4LST}
            Expected error in the measurement of the diameter of a star with the diameter and declination of $\gamma$~Crv as a function of its magnitude for the two MAGIC telescope interferometer, a second interferometer consisting of the two MAGIC telescopes and the already existing LST-1 and a third interferometer incorporating the two MAGIC telescopes and the four LSTs that will be built at ORM in La Palma. The simulated total observing time across a single night is 140 minutes. Error bars are the resulting 1-sigma dispersion of 100 simulated iterations, while the dashed lines show the linear trend of the relative error that would be achieved under dark-sky conditions (negligible night-sky background contribution).
          }
\end{figure}
The primary mirrors of the CTA IACTs will be of three different sizes, optimized for different $\gamma$-ray energy ranges. The proposed sub-arrays of four Large-Sized Telescopes (LST\cite{lst}) at CTA-North and CTA-South should be equipped with the largest mirror area (roughly 400~m$^2$).
The team within CTA responsible for the development of the LST is the LST international collaboration. This collaboration completed the design of the telescope in 2015 and finished the hardware installation of a full-size prototype (LST-1) at the CTA-North site at ORM in 2018, at a distance of around 100~m from the MAGIC telescopes. It is completing its software integration and scientific validation. The LST collaboration has essentially finished the production of the elements for the next three LSTs at ORM and installation is scheduled during the next 2 years.

We have made a simulation to understand the performance of three different interferometers: the current MAGIC interferometer, a setup including MAGIC and LST-1, and a third one with MAGIC and the four LSTs in CTA-North. Figure \ref{fig:error-diameter-MAGIC-4LST} shows the expected error in the measurement of the diameter of a star with the diameter and declination of $\gamma$~Crv (respectively 0.72~mas and  -17.5$^{\circ}$) but different magnitudes in the B~band. We use this star as a reference because these two parameters are well adapted to the baseline and latitude of MAGIC. The three straight lines correspond to an ideal case when the night sky background is negligible. The points correspond to a more realistic estimate including Full Moon background. The limit magnitude increases by about 2.8 mag when incorporating the LSTs into the interferometer. As a reference the MAGIC+4 LST interferometer can measure the diameter of a B=6.2$^m$ star with a precision of 10\% (in dark or low Moon conditions) in roughly 2h. But this is not the only advantage of this interferometer: with its 15 different telescope pair combinations one can measure the diameter in different orientations in the sky and simultaneously cover baselines from $\sim$20~m to 250~m.

The design of the LST was inspired by MAGIC so many of its parameters are similar. The reflector is equally tessellated (198 hexagonal 2~m$^2$ mirror tiles) and parabolic in shape but has a larger diameter, 23~m, and a larger f/D=1.2. The telescope is also equipped with an AMC. The camera has 1855 Hamamatsu PMTs with a peak quantum efficiency of 42\% (R11920-100). Like MAGIC, each pixel has a FOV of 0.1$^{\circ}$. The distance between PMTs is 5~cm. The camera has a modular design but, contrary to MAGIC, the digitization and trigger electronics are embedded inside the camera: each 7-pixel ''module'' hosts 7~PMTs, power supply, front-end electronics, digitizer and trigger electronics. The trigger signal produced by a module is distributed to its seven neighbouring modules through a PCB (dubbed ''backplane'').

The following design may serve to add the four LSTs in CTA-North to the current MAGIC interferometer:
\begin{itemize}
\item Same as in MAGIC we will benefit from the mechanical structure that is used to deploy the white target in front of the PMTs to bring interference filters in front of the PMTs. The filters are of the same company and model but are larger in size (75 mm).
\item In contrast to MAGIC, the filters can be positioned in front of any pixel of the camera. This actually allows us to select pixels near the center of the camera for which the PSF and time distribution are optimal.
\item Same as in MAGIC or most of the IACTs, the reflector is not focused to stars at infinity but the AMC software also allows to change focus.
\item The main challenge is the fact that the PMT signals are not delivered to a central location outside the telescope, as is the case in MAGIC, so we need to add newly designed electronics to the current camera. We plan to specifically extract one of the trigger signals from the trigger mezzanine card (currently not used) and drive that signal over the backplane to a VCSEL similar to the one used in MAGIC.
\item The optical signal generated by the VCSEL will be sent over an optical fiber to the MAGIC counting house.
\item The MAGIC custom-made optical receiver and amplifier will be used to deliver the signal to a Spectrum digitizer.
\item We expect the correlator described in the previous section to be able to handle the signals produced by the two MAGIC and four LST telescopes at ORM.
\end{itemize}

The design to extract the signal from the PMTs and generate the optical signal has been tested in the lab and has a minimal effect in the VHE acquisition chain. We are manufacturing a more compact and integrated PCB. This PCB will be tested once again at the lab. If its performance is satisfactory, we are planning a field test in LST-1. Two multimode optical fibers are already deployed from the camera of LST-1 to the MAGIC counting house. As a first test we plan to use the current correlator, which is able to process up to four channels.

\subsection{A faster digitizer in MAGIC}

High speed acquisitions, up to 4GHz, are becoming possible using modern FPGAs.
%
Such hardware is pushed forward by the advent of 5G telephony, and the industry created devices with
a monolithic integration of ADCs within a large FPGA fabric and ARM processors, creating a complete SOC (System On a Chip) aimed at processing Radio frequency signals. Such approach has many advantages, and in particular it eliminates the bottleneck of transporting the data from the ADCs to the processing system.

\begin{figure} [ht]
  \begin{center}
    \begin{tabular}{c} 
      \includegraphics[height=6cm]{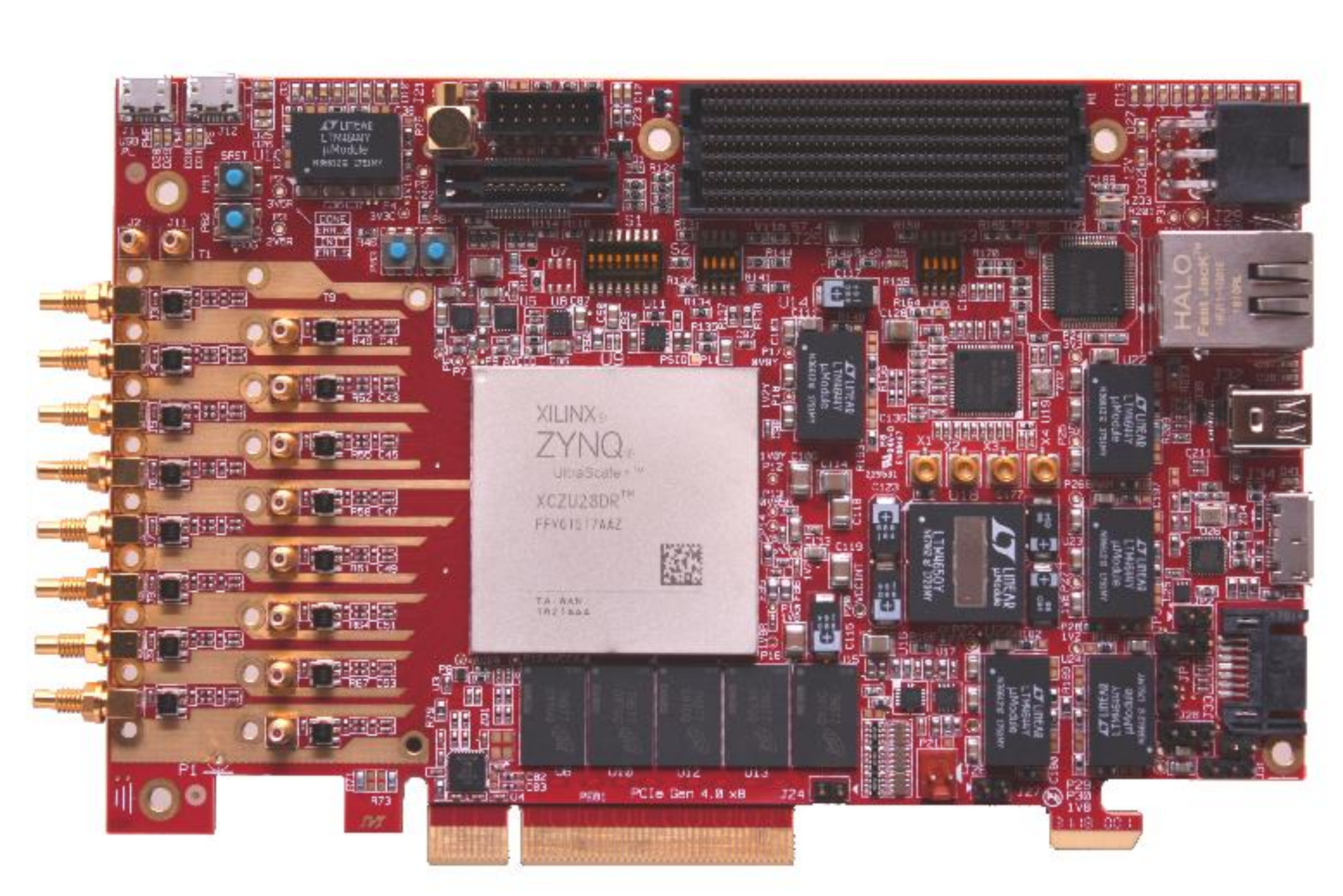}
    \end{tabular}
  \end{center}
  \caption[htg]
          { \label{fig:htg}
            A picture of the HTG-ZRF8 card under test in MAGIC.
          }
\end{figure}
We selected the HTG-ZRF8 from HiTech Global to conduct a first evaluation of this technology applied to SII. This card is populated with one Xilinx ZYNQ UltraScale+ RFSoC ZU28DR, and features 8 ADCs with a resolution of 12 bits at up to 4GHz sampling rate per ADC. The FPGA itself is very capable, with 930k cells, 4272 Digital Signal Processors (DSPs) and 4 ARM CPU cores.

The signal from 2 pixels from each MAGIC telescopes is carried via optical fiber up to the MAGIC counting house, where it is converted to an electric signal. We share the fibers and converter boards with the classical GPU-based setup, and use a dedicated line only at the output of the optical-to-electrical converter. This allows to exploit both acquisition systems simultaneously and haven't registered any cross-system interference yet. Our acquisition system is connected to the lower four inputs of the HTG board.

Due to the somewhat more limited resources compared to a GPU, the computation of the correlation between input channels is implemented in the time domain, without going through the computation of an FFT. This limits the width of the correlation window to a few hundred samples at most.

Right after sampling, the correlation begins by summing the samples and their squares using DSPs. The duration of a given summation is limited by the resolution of DSPs which is 48 bits, hence a maximum of $2^{24}$ samples can have their squares summed before an overflow of the DSPs becomes likely. This requires that accumulators are readout and reset at 239Hz assuming a 4GHz sampling rate. In reality, the card uses 8 parallel lanes to output the ADC values. Our implementation uses the same approach and 8 DSPs are used in parallel for each ADC summation. As we compute 6 correlations from 4 channels, a total of 64 DSPs are required for the accumulations alone. We calculate the mean and sigma of each accumulation window from these partial sums in the ARM cores as it requires a square root and occurs at a lower rate.

Correlations are calculated for 6 pairs of channels and for a fixed number of delays. This uses 48 DSPs, while delays are implemented using variable-depth FIFOs. Currently the number of delays is set to 80, and is inherently limited by the number of DSPs available on the FPGA. FIFOs depth can be set at runtime to change the delay applied to each channel. This is required to compensate for parallax effects and differences in fibers length.

The system currently runs at 2GHz, with correlation windows of 1 millisecond. One ARM core collects accumulated data at 1~kHz from the FPGA and pushes it to an intermediate storage using the Ethernet link of the card. The ARM core also monitors status registers of the ADCs and handle overflows by restarting the acquisition. All overflows are registered in the acquisition chain for off-line handling and quality monitoring. This type of problem remains sparse though, and the up-time of the system is in the order of 99.99\%\\

The system was tested at the MAGIC site last February and created no disturbance to the regular observations. It is now fully integrated in the regular observations execution and under active development and firmware debugging.

\subsection{Acknowledgments}
The MAGIC and LST Collaborations would like to thank the Instituto de Astrof\'{\i}sica de Canarias for the excellent working conditions at the Observatorio del Roque de los Muchachos in La Palma. The authors wish to acknowledge fruitful discussions with colleagues of the CTA consortium.
The financial support of the German BMBF, MPG and HGF; the Italian INFN and INAF; the Swiss National Fund SNF; the ERDF under the Spanish Ministerio de Ciencia e Innovaci\'on (MICINN) (PID2019-104114RB-C31, PID2019-104114RB-C32, PID2019-104114RB-C33, PID2019-105510GB-C31,PID2019-107847RB-C41, PID2019-107847RB-C42, PID2019-107847RB-C44, PID2019-107988GB-C22); the Indian Department of Atomic Energy; the Japanese ICRR, the University of Tokyo, JSPS, and MEXT; the Bulgarian Ministry of Education and Science, National RI Roadmap Project DO1-400/18.12.2020 and the Academy of Finland grant nr. 320045 is gratefully acknowledged. This work was also supported by the Spanish Centro de Excelencia ``Severo Ochoa'' (SEV-2016-0588, SEV-2017-0709, CEX2019-000920-S), the Unidad de Excelencia ``Mar\'{\i}a de Maeztu'' (CEX2019-000918-M, MDM-2015-0509-18-2) and by the CERCA program of the Generalitat de Catalunya; by the Croatian Science Foundation (HrZZ) Project IP-2016-06-9782 and the University of Rijeka Project uniri-prirod-18-48; by the DFG Collaborative Research Centers SFB823/C4 and SFB876/C3; the Polish National Research Centre grant UMO-2016/22/M/ST9/00382; and by the Brazilian MCTIC, CNPq and FAPERJ. T.~H. is funded with a grant of the ``Atracci\'on de Talento'' excellence program of Madrid, Spain, and M.~F. in part by the University of Padova (Italy) by the ''BIRD NALE$\_$SID19$\_$01'' project.

\bibliography{report} 
\bibliographystyle{spiebib} 

\end{document}